\documentclass[prl,noshowpacs,english,superscriptaddress,twocolumn]{revtex4}
\usepackage{amsmath,amssymb,amsfonts,mathrsfs}
\usepackage{amsthm}
\usepackage{CJK}
\usepackage{bm}
\usepackage{babel}
\usepackage{graphicx}
\allowdisplaybreaks
\usepackage{braket}
\usepackage[breaklinks]{hyperref}
\usepackage{color}
\hypersetup{colorlinks=true, linkcolor=blue, citecolor=blue, filecolor=blue, urlcolor=blue}

\begin{document}

\title{Ideal Intersecting Nodal Ring Phonons in a Body-Centered Cubic C$_{8}$}

\author{Y. J. Jin}
\affiliation{Department of Physics $\&$ Institute for Quantum Science and Engineering, Southern University of Science and Technology,
Shenzhen 518055, P. R. China.}
\author{Z. J. Chen}
\affiliation{Department of Physics $\&$ Institute for Quantum Science and Engineering, Southern University of Science and Technology,
Shenzhen 518055, P. R. China.}
\affiliation{Department of Physics, South China University of Technology, Guangzhou 510640, P. R. China}
\author{B. W. Xia}
\affiliation{Department of Physics $\&$ Institute for Quantum Science and Engineering, Southern University of Science and Technology,
Shenzhen 518055, P. R. China.}
\author{Y. J. Zhao}
\affiliation{Department of Physics, South China University of Technology, Guangzhou 510640, P. R. China}
\author{R. Wang}
\email[]{rcwang@cqu.edu.cn}
\affiliation{Department of Physics $\&$ Institute for Quantum Science and Engineering, Southern University of Science and Technology,
Shenzhen 518055, P. R. China.}
\affiliation{Institute for Structure and Function $\&$
Department of physics, Chongqing University, Chongqing 400044, P. R. China.}
\author{H. Xu}
\email[]{xuh@sustc.edu.cn}
\affiliation{Department of Physics $\&$ Institute for Quantum Science and Engineering, Southern University of Science and Technology,
Shenzhen 518055, P. R. China.}

\begin{abstract}
Carbon, a basic versatile element in our universe, exhibits rich varieties of allotropic phases, most of which possess promising nontrivial topological fermions. In this work, we identify a distinct topological phonon phase in a realistic carbon allotrope with a body-centered cubic structure, termed bcc-C$_{8}$. We show by symmetry arguments and effective model analysis that there are three intersecting phonon nodal rings perpendicular to each other in different planes. The intersecting phonon nodal rings are protected by time-reversal and inversion symmetries, which quantize the corresponding Berry phase into integer multiples of $\pi$. Unlike the electron systems, the phonon nodal rings in bcc-C$_{8}$ are guaranteed to remain gapless due to the lack of spin-orbital coupling. The nearly flat drumhead surface states projected on semi-infinite (001) and (110) surfaces of bcc-C$_{8}$ are clearly visible. Our findings not only discover promising nodal ring phonons in a carbon allotrope, but also provide emergent avenues for exploring topological phonons beyond fermionic electrons in carbon-allotropic structures with attractive features.
\end{abstract}

\pacs{73.20.At, 71.55.Ak, 74.43.-f}

\keywords{ }

\maketitle

Carbon is an extremely capable element since it is able to form a vast number of allotropes with fascinating properties. The prominent members include graphite, diamond,
carbon nanotubes \cite{carbontube}, graphene \cite{RevModPhys.81.109}, and fullerenes. Among these allotropic phases, one of the most important members is graphene, a single atomic layer of carbon atoms in a honeycomb lattice. On the one hand, the discovery of graphene has triggered tremendous interest in two dimensional (2D) materials \cite{PhysRevLett.105.136805, PhysRevLett.108.155501, NatRM}. On the other hand, more importantly, graphene significantly promotes advancements in studies of topological materials. As is well known, graphene possesses the unique electronic structure that exhibits topological semimetallic features with massless Dirac fermions in the absence of spin-orbital coupling (SOC) \cite{PhysRevLett.99.236809}. Furthermore, when the SOC effect is considered, the Dirac cone is destroyed and then graphene converts into a quantum spin Hall (QSH) insulator (i.e., 2D topological insulator) \cite{PhysRevLett.95.226801}. Topological insulators \cite{Kane-RevModPhys.82.3045, ZSC-RevModPhys.83.1057} and topological semimetals \cite{RevModPhys.90.015001} have attracted extensive attentions over the past decade.
Beyond 2D graphene, nontrivial electronic states have also been intensively extended into three dimensional (3D) carbon allotropes. For instance, the topological semimetallic phases have been predicted in 3D graphene networks \cite{PhysRevB.92.045108, nanolett2015dd, PhysRevB.97.245147}, body-centered orthorhombic C$_{16}$ \cite{PhysRevLett.116.195501}, body-centered tetragonal C$_{16}$ \cite{smallC16BCT} and C$_{40}$ \cite{PhysRevLett.120.026402}, as well as even beyond \cite{C12SR, FENG2018527, Carbonnc, CHENG2016468, C6NR00882H, PhysRevLett.115.026403}.

Overall, these advances mentioned above provide exciting avenues for exploring band topology in carbon allotropes. Motivated by observations of topological fermions, we naturally raise a question: are there carbon allotropes that can host topological bosons?
In fact, besides nontrivial fermionic electrons, the topological states have recently been accessed into bosonic systems \cite{PhysRevLett.100.013904, PhysRevLett.100.013905, Natcom2015, Natphys2016, Science2015, Natphys2018, PhysRevB.97.035442, PhysRevLett.120.016401, PhysRevLett.121.035302, PhysRevB.97.054305, mgbphonon}. In the subjects of nontrivial bosons, topological phonons (i.e., quantized excited vibrational states of interacting atoms) are particular importance, which provide potentially promising applications of electron-phonon coupling, dynamic instability \cite{PhysRevLett.103.248101}, and  phonon diode \cite {PhysRevB.96.064106}. Very recently, topological phonons at THz frequencies have been proposed in a few materials, such as double Weyl phonons in transition-metal monosilicides \cite{PhysRevLett.120.016401, PhysRevLett.121.035302}, Weyl and triple phonons in WC-type compounds \cite{PhysRevB.97.054305}, and Weyl nodal straight line phonons in MgB$_{2}$ \cite{mgbphonon}. However, the realistic candidates with topological phonons discovered to date are very limited. As the topological orders in carbon allotropic materials host the most attractive features, it is highly desirable to explore topological phonon states in carbon allotropes.

In this work, using first-principles calculations and topological analysis, we identify that a carbon allotrope with a body-centered cubic (bcc) structure in space group $Im$-$3m$ (No. 229) exhibits exotic topological phonon states. This carbon phase contains eight atoms in one primitive unit cell, thus termed bcc-C$_{8}$. We show by symmetry arguments that bcc-C$_{8}$ hosts three phonon nodal rings in different planes intersecting at six different points. We reveal the quantized Berry phase and drumhead surface states to further support its topological phonon features. Importantly,  bcc-C$_{8}$ has already been synthesized in experiments \cite{C8, CHEN2014165}, suggesting that it is of fundamental importance and practical interests to establish a realistic carbon phase with intriguing nontrivial phonons.

We performed first-principles calculations based on density functional theory \cite{Kohn} as implemented in the Vienna \textit{ab initio} simulation package \cite{Kresse2}. The generalized gradient approximation with Perdew-Burke-Ernzerhof functional was employed to describe exchange-correlation energy. The core-valence interactions were treated by the projector augmented wave method  \cite{Kresse4,Ceperley1980}. The cutoff energy of plane-wave was taken as 500 eV, and the Brillouin zone (BZ) was sampled by $12\times12\times12$ Monkhorst-Pack grid \cite{Monkhorst}. The structural parameters of bcc-C$_{8}$ were optimized by minimizing the forces on each atom smaller than $1.0\times10^{-3}$ eV/{\AA}. We calculated lattice dynamics using density-functional perturbation theory (DFPT) \cite{RevModPhys.73.515} with a $3\times3\times3$ supercell. The phonon spectra were obtained by diagonalization of real-space force constants as implemented in the PHONOPY package \cite{TOGO20151}. In order to reveal nontrivial features of phonons in bcc-C$_{8}$, we also constructed a Wannier tight-binding (TB) Hamiltonian of phonons from the second rank tensor of force constants \cite{TOGO20151}.

\begin{figure}
	\centering
	\includegraphics[scale=0.39]{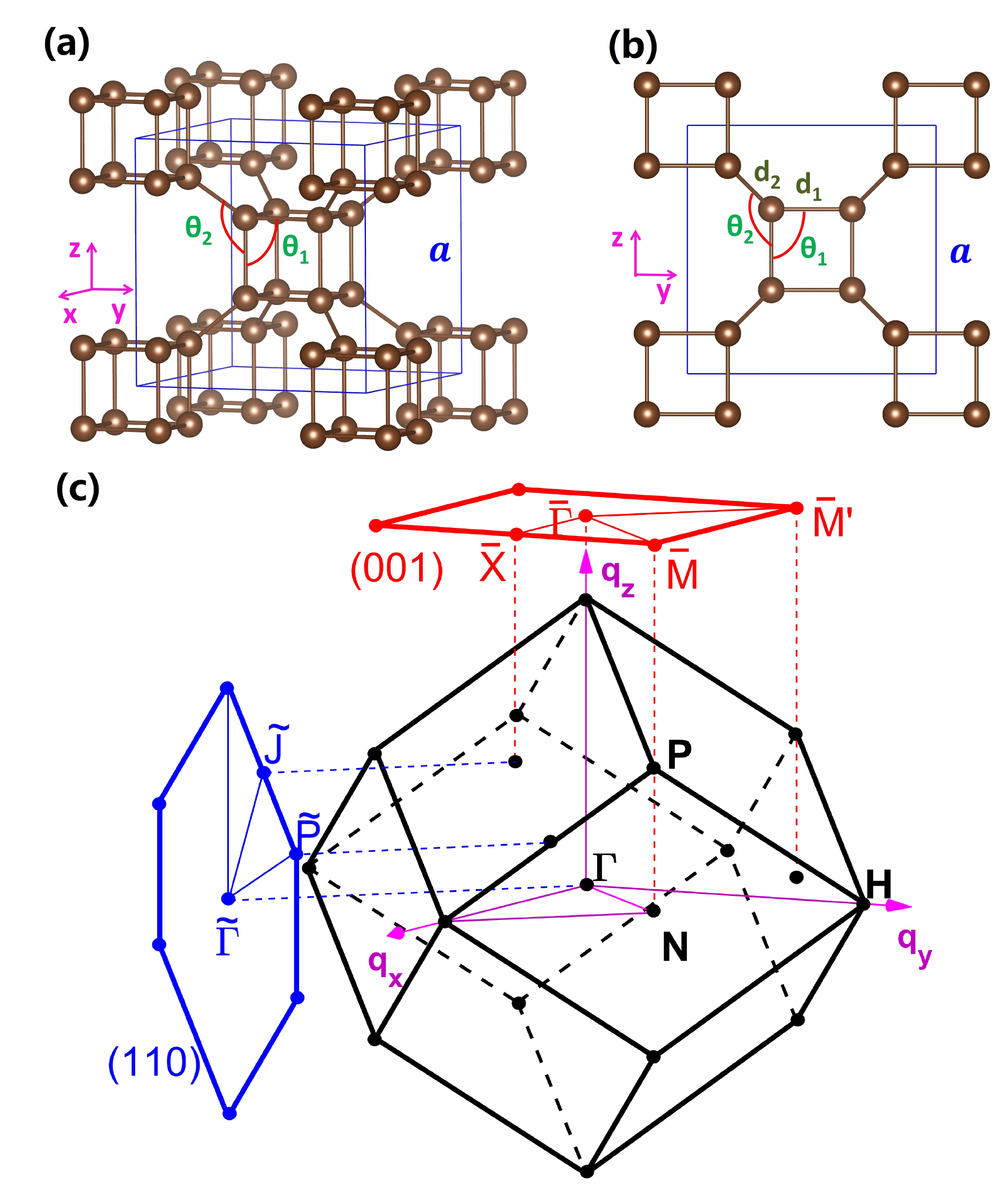}
	\caption{(a) Side and (b) top views of bcc-C$_8$ crystallized in space group \emph{Im-3m} (No. 229). The two distinct bond lengths ($d_1$, $d_2$) and bond angles ($\theta_1$, $\theta_2$) are denoted. (c) The bulk BZ and its corresponding surface BZ projected on the (001) and (110) surfaces, respectively.
\label{figure1-struct-bz}}
\end{figure}

As shown in Figs. \ref{figure1-struct-bz}(a) and \ref{figure1-struct-bz}(b), we can see that each lattice point in bcc-C$_8$ consists of eight carbon atoms forming a sub-cube, and each carbon atom bonds to four neighbors showing a distorted tetrahedron. The eight carbon atoms in one primitive unit cell occupy an equivalent Wyckoff position of 16f (-0.3374, -0.3374, 0.3374).  Its optimized lattice constant is $a=4.87$ {\AA}, which is in good agreement with the experimental value $a=4.84$ {\AA} \cite{CHEN2014165}. In contrast to the unique bond-length in 2D graphene or 3D diamond, there are two distinct carbon-carbon bond lengths. The longer bond $d_1 = 1.586$ {\AA} is associated with intra-bonds in a sub-cube, and the shorter one $d_2 = 1.476$ {\AA} connects two neighboring sub-cubes. There are also two different bond angles, i.e., $\theta_1 = 90^{\circ}$ in a sub-cube and $\theta_2 = 125.26^{\circ}$ out of a sub-cube. In Fig. \ref{figure1-struct-bz}(c), we show the bulk BZ and the (001) and (110) surface BZs, in which high-symmetry points are marked.

\begin{figure}
\setlength{\abovecaptionskip}{-0.10cm}
	\centering
	\includegraphics[scale=0.375]{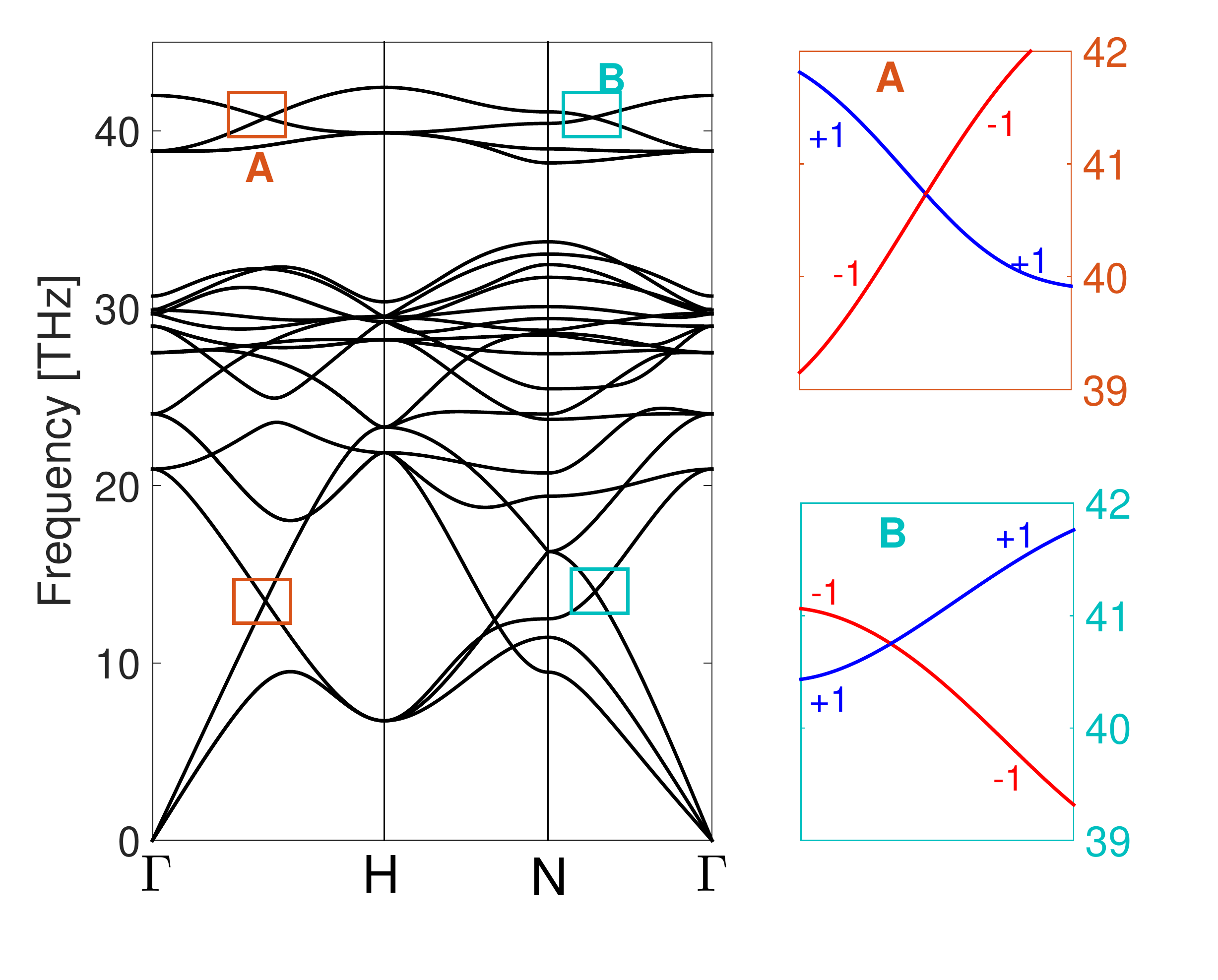}
	\caption{The phonon spectra of bcc-C$_8$ along high-symmetry lines. The right panels show the enlarged regions A and B, respectively. Two crossing branches in the $\Gamma$-$H$ and $\Gamma$-$N$ directions are classified by opposite mirror eigenvalues $\pm 1$.
\label{band-phonon}}
\end{figure}

\begin{figure}
	\centering
	\includegraphics[scale=0.3]{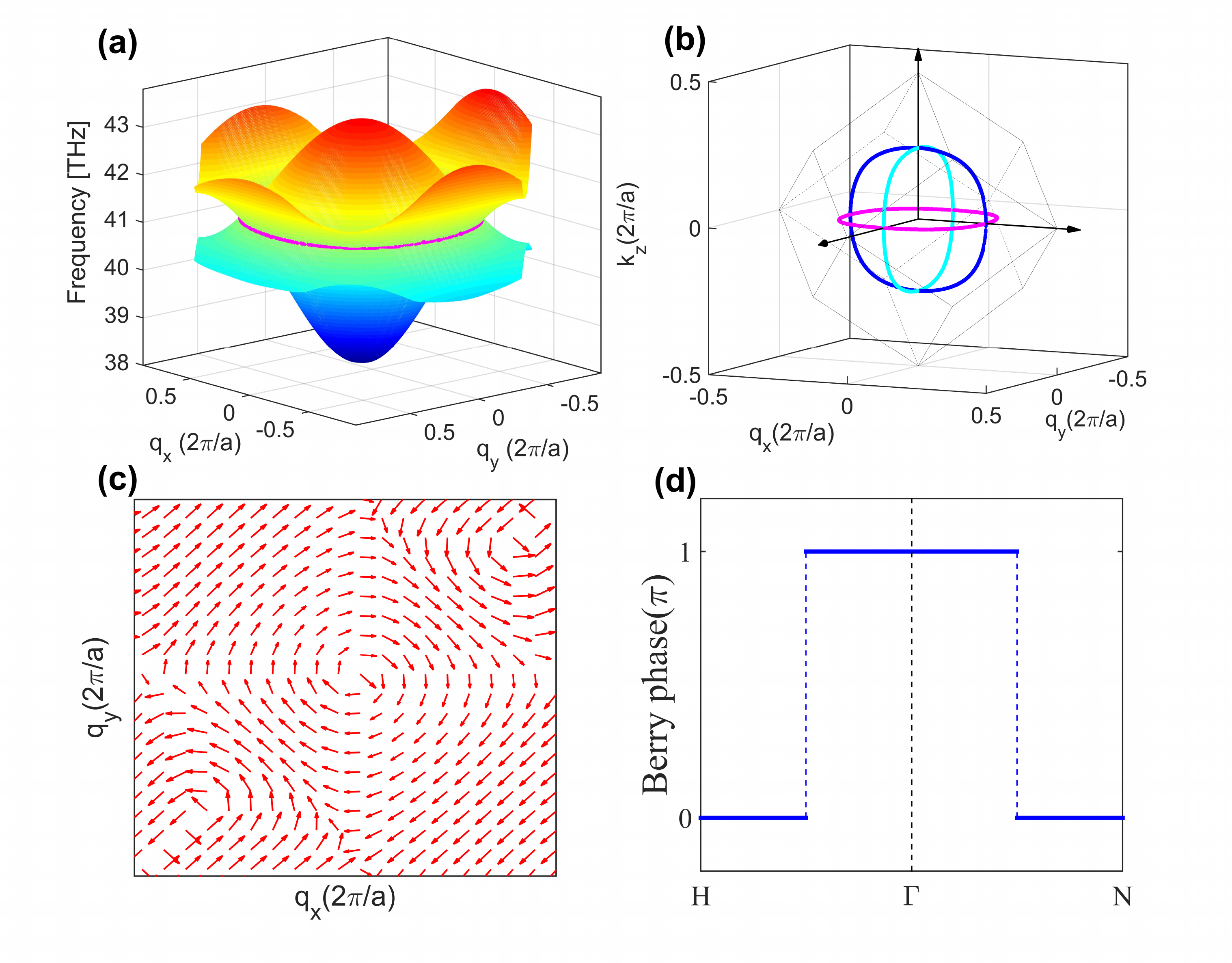}
	\caption{(a) Two crossing  branches optical phonons forming a nodal ring (colored by pink) in the $q_x$-$q_y$ plane with $q_z = 0$. (b) Three intersecting phonon nodal rings (colored by pink, blue, and light-blue, respectively) in three different planes in 3D momentum space. (c) Berry curvature distribution of a nodal ring in the $q_x$-$q_y$ plane with $k_z = 0$. (d) A variation of Berry phase along the high-symmetry paths $H$-$\Gamma$-$N$ in the $q_x$-$q_y$ plane with $q_z = 0$.
\label{berry}}
\end{figure}

The phonon spectra of bcc-C$_8$ along the high-symmetry paths of the BZ are plotted in Fig. \ref{band-phonon}(a). It is clear to see that there are two sets of visible double-degenerate points in both the $\Gamma$-$H$ and $\Gamma$-$N$ directions. One comes from the crossing of the longitudinal acoustic branch and the lowest transverse optical branch, and the other is originated from the crossing of two highest optical branches. As phonons are typically bosons, they are not limited by the Pauli exclusion principle. That is to say, the topological features in the whole range of phonon frequencies can be detected. Therefore, we here only focus on the crossings of two highest optical branches, for which the two nontrivial crossing branches are well separated from trivial ones. As shown in the right panels of Fig. \ref{band-phonon}, the enlarged views of phonon spectra show the crossing points in the $\Gamma$-$H$ and $\Gamma$-$N$ directions at the same frequency of $\omega_D = 40.75$ THz. Actually, these crossings with linear dispersions are in the $q_x$-$q_y$ plane with $q_z = 0$, which is with respect to the mirror reflection symmetry $M_z$. The two crossing branches within this mirror-reflection invariant plane belong to two opposite mirror eigenvalues $\pm 1$. A 3D plot of these crossing branches with $q_z =0$ is present in Fig. \ref{berry}(a). The figure shows that the crossings of two inverted branches form a continuous phonon nodal-ring. The nodal ring exhibits no frequency dispersion in the $q_x$-$q_y$ plane. This ideal feature of topological phonons in bcc-C$_8$ can be easily detected and is important to applications of topological quantum transport of phonons. Besides the phonon nodal ring in the $q_x$-$q_y$ plane, we also find another two phonon nodal rings in the $q_x$-$q_z$ plane and $q_y$-$q_z$ plane, respectively. As depicted in Fig. \ref{berry}(b), the three phonon nodal rings perpendicular to each other in different planes intersect at six points.

Next, we prove the existence of three intersecting phonon nodal rings encircling the $\Gamma$ point by an effective $\mathbf{k}\cdot \mathbf{p}$ model. The symmetry at the $\Gamma$ point of bcc-C$_8$ is characterized by $O_h$, which includes inversion symmetry $\mathcal{I}$, 4-fold rotational symmetries $C_4$ around the $x$, $y$, and $z$ axes, and mirror-reflection symmetries $M_x$,  $M_y$, and  $M_z$.  In general, the two crossing branches of phonons can be described by a $2\times 2$ low-energy $\mathbf{k}\cdot \mathbf{p}$ Hamiltonian as
\begin{equation}\label{Hamiltonian}
\mathcal{H}(\mathbf{q})=\sum_{i=x,y,z}d_i(\mathbf{q})\sigma_i,
\end{equation}
where $\mathcal{H}$ is referenced to the frequency of a crossing point, $d_i(\mathbf{q})$ are real functions, $\mathbf{q}=(q_x, q_y, q_z)$ is the phonon wavevector, and $\sigma_i$ are three Pauli matrices. The identify matrix $\sigma_0$ only shifts degenerate points of phonons and can be ignored in Eq. (\ref{Hamiltonian}) in the following. The two crossing branches of phonons around the $\Gamma$ point have the opposite eigenvalues, indicating that the $\mathcal{I}$ symmetry can be chosen as $\mathcal{I}=\sigma_z$. The Hamiltonian has the following constraint as
\begin{equation}
\mathcal{I}\mathcal{H}(\mathbf{q}) \mathcal{I}^{-1} =  \mathcal{H}(-\mathbf{q}),
\end{equation}
which forces that $d_{x,y}(\mathbf{q})$ are odd functions and $d_z(\mathbf{q})$ is an even function. Besides, phonons are typically spinless \cite{Levine1962}, and thus the time-reversal ($\mathcal{T}$) symmetry of a phonon system is always conserved when there is no any strain-gradient field. The absence of SOC indicates $\mathcal{T}^2=1$; that is, the $\mathcal{T}$ symmetry can be represented by $\mathcal{T}=K$, where $K$ is the complex conjugate operator. The $\mathcal{T}$ symmetry requires
\begin{equation}
\mathcal{T}\mathcal{H}(\mathbf{q}) \mathcal{T}^{-1} =  \mathcal{H}(-\mathbf{q}),
\end{equation}
which gives a constraint leading to that $d_{x,z}(\mathbf{q})$ are even functions and $d_y(\mathbf{q})$ is an odd function. As a result, the coexistence of $\mathcal{I}$ and $\mathcal{T}$ symmetries gives $d_x(\mathbf{q})\equiv 0$, $d_y(\mathbf{q})=-d_y(\mathbf{-q})$, and $d_z(\mathbf{q})=d_z(\mathbf{-q})$. The symmetry-allowed expressions as a function of $\mathbf{q}$ in the low energy can be generally written as
\begin{equation}\label{expression}
\begin{split}
&d_y(\mathbf{q})=\sum_{i=x,y,z}C_{1y}^i q_i +\sum_{i,j,k=x,y,z} C_{3y}^{ijk}q_i q_j q_k,\\
&d_z(\mathbf{q})= C_{0z}+\sum_{i,j=x,y,z}C_{2z}^{ij}q_i q_j.
\end{split}
\end{equation}
The generic solutions of two phonon crossing branches require $d_y(\mathbf{q}) =0$ and $d_z(\mathbf{q}) =0$. Based on Eq. (\ref{expression}), this condition has codimension one, allowing nodal-lines in momentum space. The rotational symmetries $C_4^{i}$ ($i=x,y,z$) and mirror symmetries $M_i$ introduce additional symmetry constraints on $d_y(\mathbf{q}$) and $d_z(\mathbf{q})$. Under these symmetries, Eq. (\ref{expression}) can be reduced as
\begin{equation}\label{expreduced}
\begin{split}
&d_y(\mathbf{q})= C_{3y}^{xyz}q_x q_y q_z,\\
&d_z(\mathbf{q})= C_{0z}+C_{2z}^{xx}(q_x^2+q_y^2+q_z^2),
\end{split}
\end{equation}
where the condition of band inversion requires $C_{0z}C_{2z}^{xx}<0$. In this case, Eq. (\ref{expreduced}) indicates the appearance of three intersecting closed nodal rings perpendicular to each other in the $q_x$, $q_y$, and $q_z=0$ planes. These planes are mirror-reflection invariant. However, it is worth noting that the stability of phonon nodal rings in bcc-C$_8$ is topologically protected by the coexistence of $\mathcal{I}$ and $\mathcal{T}$ symmetries. The additional rotational and mirror symmetries just force the phonon nodal rings to be related to $C_4$ symmetries around the $q_x$, $q_y$, and $q_y$ axes and lie in the $q_x$, $q_y$, and $q_z=0$ planes, respectively. Due to the lack of the SOC effect in a phonon system, the phonon nodal rings in bcc-C$_8$ are generally robust and guaranteed to be gapless with respect to perturbations as long as $\mathcal{I}$ and $\mathcal{T}$ symmetries are present.

\begin{figure}
	\centering
	\includegraphics[scale=0.4]{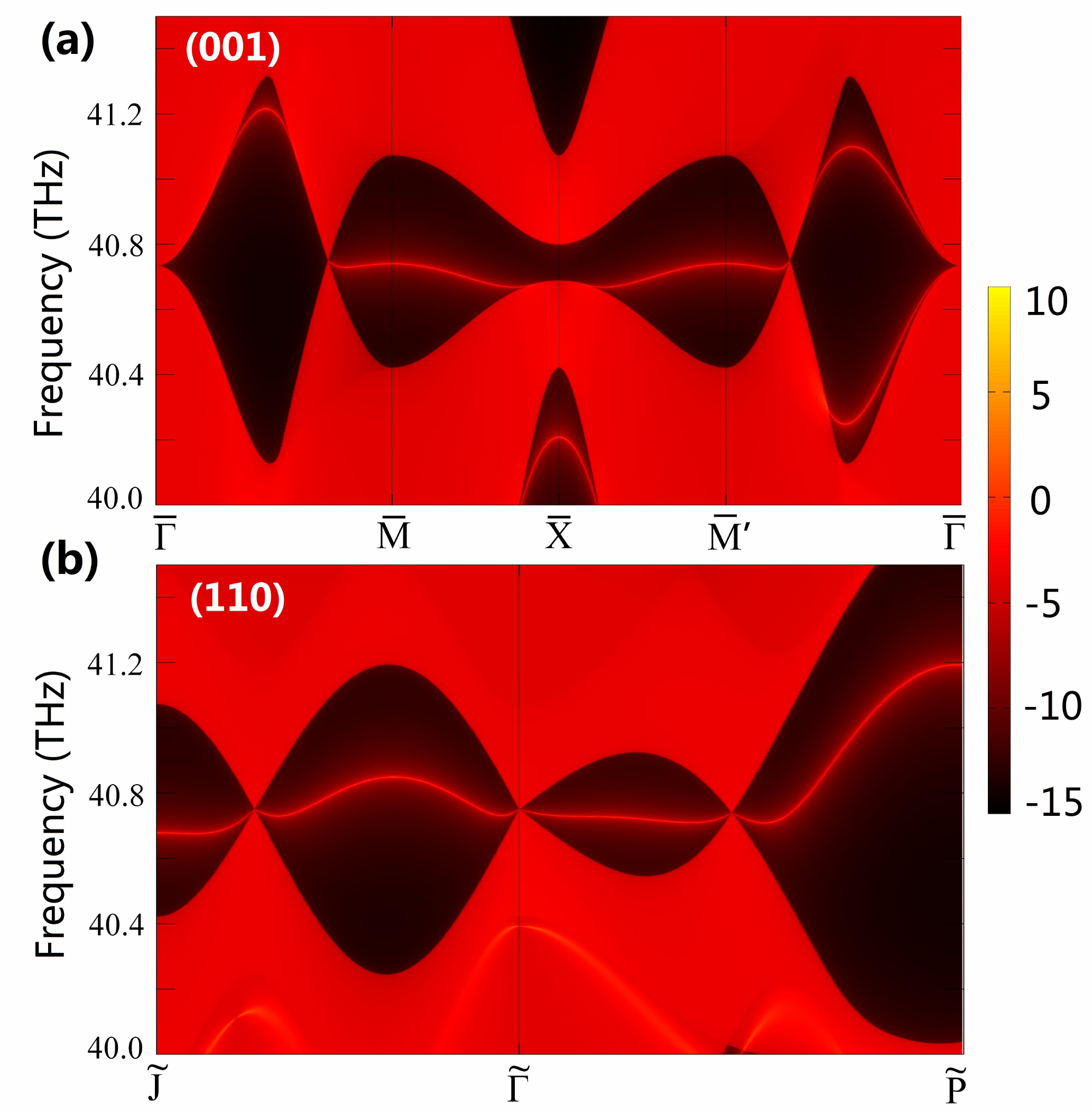}
	\caption{ The phonon surface states of bcc-C$_8$. (a) LDOS projected on the semi-infinite (001) surface. (b) LDOS projected on the semi-infinite (001) surface. Red regions represent the projections of bulk phonon branches, and red lines represent the phonon surface states. The nontrivial drumhead surface states terminated at the projections of phonon crossing points are nearly flat and clearly visible.
\label{surface}}
\end{figure}

The presence of topological phonon nodal rings corresponds to Berry phase quantization \cite{PhysRevB.92.081201}. The Berry phase of a closed loop $\mathcal{C}$ in 3D momentum space is defined as
\begin{equation}
\gamma=\oint_\mathcal{C} \mathcal{A}(\mathbf{q})\cdot d\mathbf{q},
\end{equation}
where $\mathcal{A}(\mathbf{q})=-i\sum_{\lambda}{\langle \varphi_{\lambda}(\mathbf{q})|\nabla_{\mathbf{q}}|\varphi_{\lambda}(\mathbf{q})\rangle}$ is the Berry connection and $\varphi_{\lambda}(\mathbf{q})$ is the  Bloch wavefunction of the $\lambda$th phonon branch. The phonon Bloch wavefunction can be written as  \cite{PhysRevB.96.064106}
 \begin{equation}\label{wavefunction}
\varphi_{\lambda}(\mathbf{q}) = \left(
                                  \begin{array}{c}
                                    \mathcal{D}_{\mathbf{q}}^{1/2} \mathbf{u}_{\mathbf{q}} \\
                                    \mathbf{\dot{u}}_{\mathbf{q}} \\
                                  \end{array}
                                \right),
\end{equation}
where $\mathcal{D}(\mathbf{q})$ is the lattice dynamic matrix, $\mathbf{u}_{\mathbf{q}}$ represents the atomic eigen-displacement, and $\mathbf{\dot{u}}_{\mathbf{q}}$ is its time derivative. Equation (\ref{wavefunction}) indicates that we are able to calculate the phonon Berry connection $\mathcal{A}(\mathbf{q})$ using atomic force constants, which can conveniently be obtained from DFPT \cite{RevModPhys.73.515}. The corresponding phonon Berry curvature is $\mathcal{B}(\mathbf{q})=\nabla \times \mathcal{A}(\mathbf{q})$. As shown in Fig. \ref{berry}(c), the distribution of Berry curvature in the $q_z = 0$ plane exhibits a nontrivial vortex feature in momentum space. To further support the topologically protected phonon nodal rings, we also calculate a variation of phonon Berry phase in the $q_z = 0$ plane, which corresponds to the one-dimensional system along the high-symmetry paths $H$-$\Gamma$-$N$. Figure \ref{berry}(d) shows a jump of $\pi$ across the nodal ring, confirming the nontrivial phonon Berry phase ($\pi$ mod $2\pi$) of the phonon nodal ring in bcc-$C_8$.

The topological phonon nodal rings with quantized Berry phase lead to the nontrivial drumhead surface states. To directly illustrate this, we calculate the local density of states (LDOS) of phonons with the iterative Green's function method based on a phonon TB Hamiltonian \cite{Sancho1984, WU2017}. In Figs. \ref{surface}(a) and \ref{surface}(b), we plot the LDOS projected on the semi-infinite (001) and (110) surfaces of bcc-C$_8$. As expected, the topological drumhead surface states terminated at the projections of phonon crossing points are nearly flat and clearly visible. On the (001) surface, the bulk Dirac cones along  $\bar{\Gamma}$-$\bar{M}$ and $\bar{\Gamma}$-$\bar{M}'$ are projected from  the nodal ring in the $q_z = 0$ plane. As the appearance of surface breaks the $C_4$ rotational symmetry, the surface states along  $\bar{\Gamma}$-$\bar{M}$ and $\bar{\Gamma}$-$\bar{M}'$ in the square (001) surface BZ are non-equivalent [see Fig. \ref{surface}(a)]. On the (110) surface, the bulk Dirac cone at the $\tilde{\Gamma}$ point is projected from two overlapping crossing points on the $q_x=-q_y$ axis across the nodal ring in the plane of  $q_z = 0$, while the bulk Dirac cone along  $\tilde{\Gamma}$-$\tilde{J}$ (or $\tilde{\Gamma}$-$\tilde{P}$) is projected from crossing points of two superimposed nodal rings lied in the planes of $q_x = 0$ and $q_y = 0$. As a result, the intersecting orthogonal nodal rings in bcc-C$_8$ derive that the bulk Dirac cones projected on the (110) surface are connected by two nontrivial surface states.

In summary, using first-principles and effective model analysis, we propose an emergent topological bosonic phase that nontrivial phonons are present in a carbon allotrope bcc-C$_{8}$. In this exotic topological phase, there are three intersecting phonon nodal rings perpendicular to each other in different planes. The symmetry arguments indicate that these intersecting phonon nodal rings are protected by time-reversal and inversion symmetries. Due to the lack of SOC in phonon systems, the phonon nodal rings in bcc-C$_{8}$ are generally guaranteed to remain intact with respect to perturbations. Our calculations confirm that the Berry phase is quantized into integer multiples of $\pi$. The nearly flat drumhead surface states projected on the semi-infinite (001) and (110) surfaces of bcc-C$_{8}$ are clearly visible, further supporting the topological phonon features. Our findings not only provide exciting avenues for exploring topological phonons beyond nontrivial fermions in carbon allotropic structures, but also establish a realistic carbon allotrope with intriguing nontrivial phonons.

~~~\\
~~~\\
\textbf{ACKNOWLEDGMENTS}\\
 This work is supported by the National Natural Science Foundation of China (NSFC, Grant Nos.11674148 and 11304403), the Guangdong Natural Science Funds for Distinguished Young Scholars (No. 2017B030306008), and the Fundamental Research Funds for the Central Universities of China (Nos. 106112017CDJXY300005 and cqu2018CDHB1B01).\\


\end{document}